\documentclass{article}
\usepackage[utf8]{inputenc}
\usepackage{authblk}
\usepackage{setspace}
\usepackage[margin=1.25in]{geometry}
\usepackage{graphicx}
\usepackage{subcaption}
\usepackage{amsmath}
\usepackage{lineno}
\usepackage{multirow}
\usepackage{chngpage}
\usepackage{float}
\usepackage{adjustbox}
\usepackage{xcolor}
\usepackage{soul}
\usepackage{cite}

\graphicspath{ {./figures/} }

\title{Direct Sensing of Remote Nuclei: Expanding the Reach of Cross-Effect Dynamic Nuclear Polarization}

\author[1]{Amaria Javed}
\author[1,2*$\dag$]{Asif Equbal}

\affil[1]{Center for Quantum and Topological Systems, New York University Abu Dhabi, United Arab Emirates.}
\affil[2]{Department of Chemistry, New York University Abu Dhabi, United Arab Emirates.}

\affil[*]{Corresponding author. Email: asif@nyu.edu}

\date{}

\begin{document}

\maketitle

\begin{abstract}
Dynamic Nuclear Polarization (DNP) has revolutionized the field of solid-state NMR spectroscopy by significantly enhancing the sensitivity of nuclear magnetic resonance experiments. Conventionally, cross effect DNP relies on biradicals to transfer polarization from coupled electron spins to nearby nuclear spins and subsequent relay to target nuclei via spin diffusion mechanism. However, the direct transfer of polarization to distant nuclei remains a significant challenge, limiting its applicability in various contexts. In this work, we propose a novel biradical design concept that involves a very strong electron-electron coupling, with a magnitude of hundreds of MHz, which enables efficient direct polarization transfer from electron spins to nuclear spins over much longer distances, exceeding 2.0 nm. We discuss the potential of this tailored biradicals in scenarios where conventional spin diffusion mechanisms are inefficient or when direct nuclear spin sensing through electron spin interactions is desired. Our study presents a promising avenue for expanding the scope of cross effect DNP in solid-state NMR spectroscopy and opens new opportunities for investigating a wide range of biological and material systems. Our research also provides insight into the DNP buildup time of commercially available biradicals.
\end{abstract}


\section{Introduction}
Dynamic Nuclear Polarization (DNP) has emerged as a powerful technique for enhancing the sensitivity of solid-state Nuclear Magnetic Resonance (NMR) spectroscopy \cite{LILLYTHANKAMONY2017120, eills2023spin}. Imagine studying a wide range of complex biological molecules, a material with unique properties, or a catalytic surface with high resolution and sensitivity at the atomic level. These are precisely the kinds of challenge where DNP opened up a treasure trove of opportunities for NMR spectroscopists \cite{jeon2019application,Tycko_Abeta,Wylie2015,Su_Griffin_2015,Long_2015,perras2015natural,tran2020dynamic,kang2019lignin, lafon2011beyond, wolf2018endogenous,lee2017interfacial}.
DNP sensitivity enhancement is achieved through the transfer of polarization from electron spins to nuclear spins through a complex interplay between electron and nuclear spins, which is controlled by microwave radiation through different mechanisms \cite{Griffin1993}. Two primary mechanisms for achieving DNP in solid-state systems are Solid Effect (SE) and Cross Effect (CE) DNP \cite{Abragam1958,jeffries1957polarization}. The SE DNP transfer mechanism entails the simultaneous flipping of both electron and nuclear spins, achieved by using off-resonant microwave irradiation \cite{hovav2010theoretical}. The intense microwave power requirement for this quantum mechanically forbidden transition poses a significant constraint, especially in the context of high-field DNP instruments \cite{GyroAmp-250GHz}.

In contrast to SE DNP, CE DNP is a more efficient process that takes advantage of electron-electron ($e$-$e$) couplings in addition to electron-nuclear ($e$-$n$) interactions \cite{hovav2012theoretical,equbal2019cross}. This is accomplished through a triple-flip transition of an electron-electron-nuclear ($e$-$e$-$n$) system, which is regulated by the strength of the $e$-$e$ coupling and $e$-$n$ hyperfine coupling. CE DNP is especially advantageous in high-magnet-field conditions, where microwave power is limited, as it requires much less microwave power than SE DNP due to the quantum mechanically allowed microwave-induced transition. The key to CE DNP's efficiency is meeting a resonance condition where the difference between the Electron Paramagnetic Resonance (EPR) frequencies ($\Delta \omega_{e}$) of the two coupled electron spins is equal to the nuclear Larmor frequency ($\omega_{0n}$).


\begin{figure}[H]
    \centering
    \includegraphics[width=1.0\textwidth]{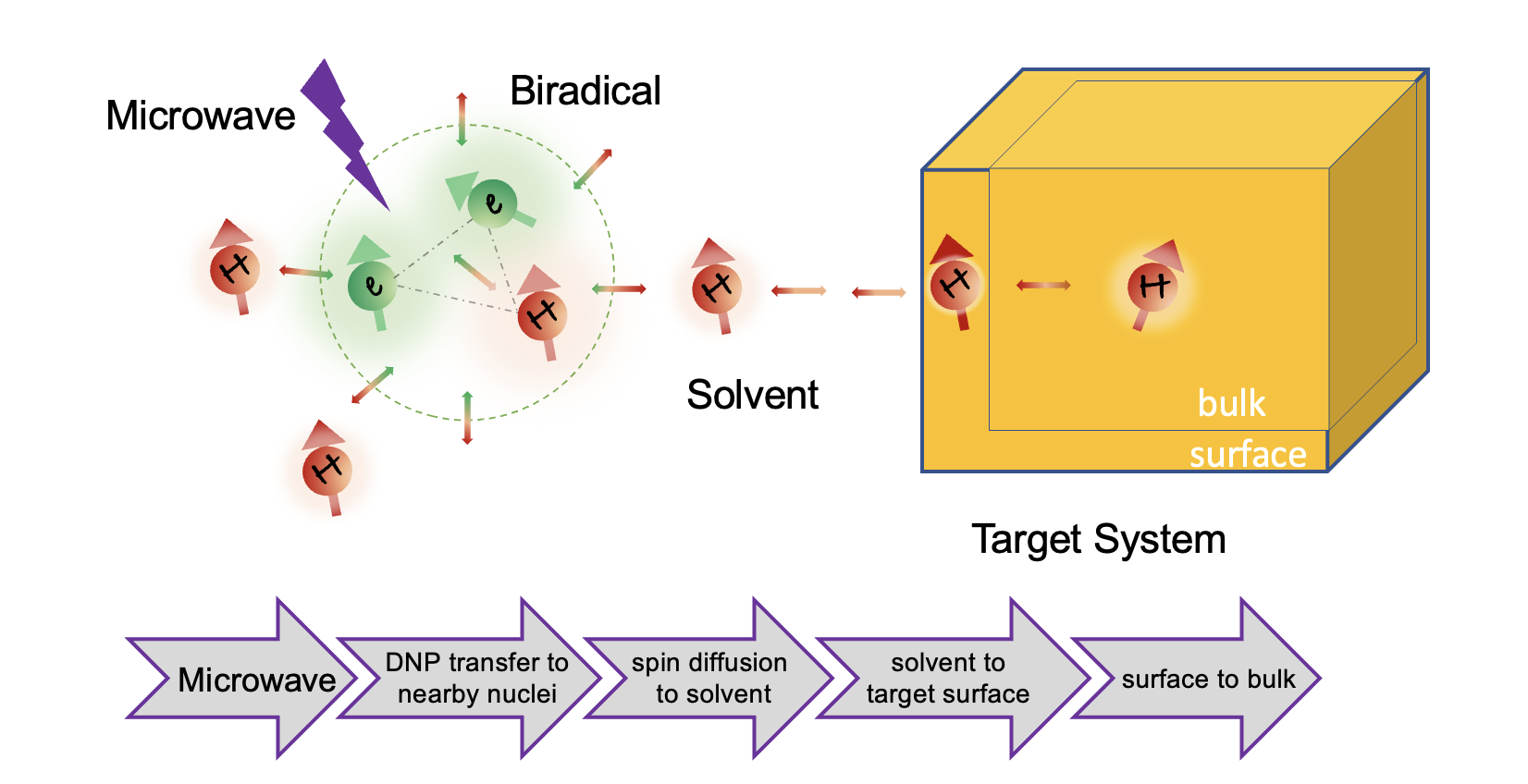}
    \caption{Schematic illustrating the conventional DNP polarization transfer mechanism, utilizing the intricate spin diffusion network within the system. This network operates within the biradical, extends into the solvent, and further propagates to the target's surface before penetrating its bulk.}
    \label{fig1}
\end{figure}

The conventional CE DNP mechanism relies on the interplay of spin properties within the polarizing agent (shown in the circle). Commonly, nitroxide-based biradicals are used in contemporary DNP applications. These biradicals have unpaired electron spins with an anisotropic g-tensor, which is larger than the nuclear Larmor frequency. At high magnetic fields, this g-anisotropy is necessary to reach the CE resonance condition, which requires the energy separation between two coupled electrons to match the nuclear Larmor frequency. Therefore, the orientation of the g-tensors of the electron spins and $e-e$ and $e-n$ couplings between the two electrons and nuclear spins must be carefully tuned. It is important to note that in conventional DNP, the polarization is transferred from electrons to the nuclei close to the electron spins. The enhanced nuclear polarization of these proximate nuclear spins (encircled) is then relayed to target nuclear spins through spin-diffusion, which involves flip-flop transitions between nuclear spins. The efficiency of DNP of the target sample process is dependent on the presence of an effective spin diffusion network connecting the polarizing agent to the target molecule.

\begin{figure}[H]
    \centering
    \includegraphics[width=1.0\textwidth]{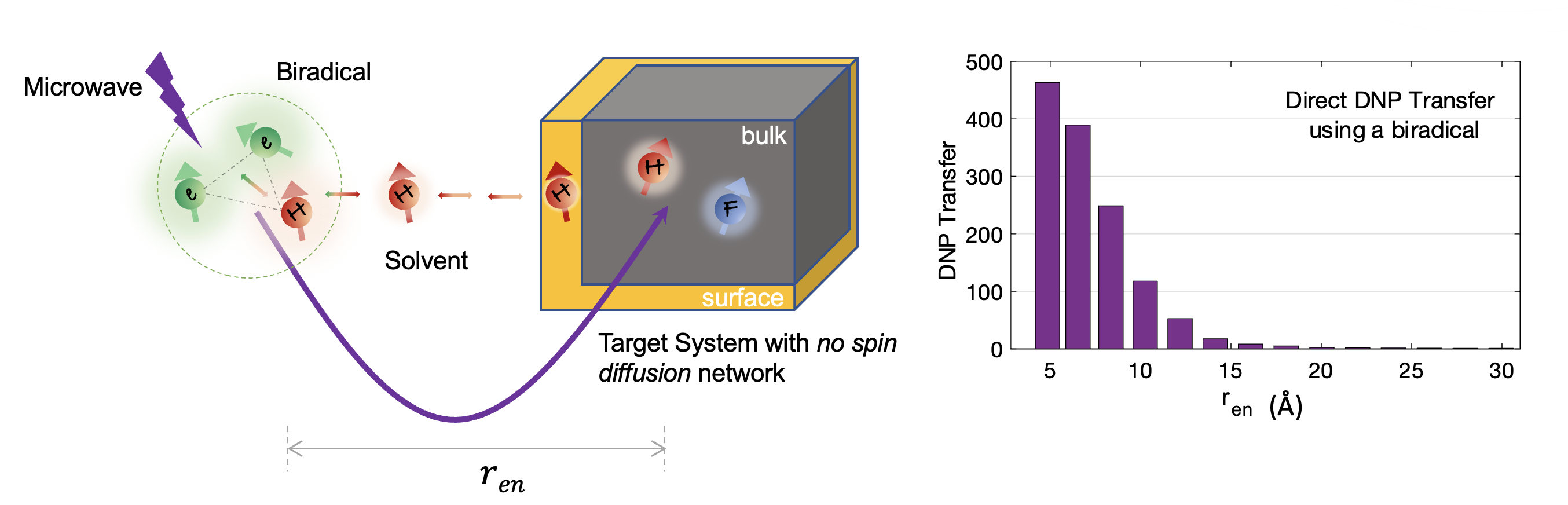}
    \caption{ Without a well-defined spin diffusion network, the enhancement of remote nuclei relies heavily on direct polarization transfer, indicated by the purple arrow. Quantum mechanical simulations of DNP reveal that using a conventional biradical ($eeH$ spin system), the direct transfer to distant nucleus ($H$) diminishes exponentially with increasing distance as shown in the right panel.}
    \label{fig2}
\end{figure}

Despite the fact that spin-diffusion based DNP transfer is useful in certain experimental setups, it has its own limitations. For example, when investigating systems with slow or negligible spin diffusion dynamics, the enhancement of a target nucleus is fundamentally dependent on the direct polarization transfer. Quantum-mechanically simulated enhancement of DNP for $H$ nuclei positioned at varying distances ($r_{en}$) from the electron spins, thereby exploring different hyperfine coupling strengths, shows that as $r_{en}$ increases, the direct CE DNP transfer exhibits a dramatic decline (Fig. \ref{fig2}). This is particularly evident in the range of small distances, with the enhancement dropping from 450 at a distance of 5 $\AA$ to a mere 120 at 10 $\AA$ at 7 T field conditions. Further separation to 20 $\AA$ results in an enhancement of less than 10, indicating a negligible polarization transfer efficiency of conventional biradicals at longer distances. These findings demonstrate a critical limitation of conventional biradicals for applications such as spin sensing or the transfer of spin polarization to nuclei (e.g. $^1H$ or $^{19}F$) situated outside the biradical molecule, to the solvent, or directly to the target molecule. Consequently, there is a need for innovative approaches that can extend the reach of CE DNP, allowing for efficient polarization transfer to nuclei at longer distances.

\section{Results}

To understand why direct DNP transfer is truncated at longer $r_{en} $distances, we need to understand the CE mechanism. In the high field approximation regime, the CE transfer Hamiltonian ($\widetilde{H}_{CE}$) for a model two electron ($S$) and a nucleus ($I$) system is dependent on the amplitude, $\omega_{een}$, and is expressed by the equation.
\begin{align}
 \nonumber \widetilde{H}_{CE} &= \omega_{CE} S_1^+S_2^-I_3^+ + c.c.\\
 \nonumber \\
 \text{where, } \omega_{CE} &= \frac{\omega_{e_1e_2}(\omega_{e_1H}-\omega_{e_2H})}{\omega_{0H}} \label{eq1}
\end{align}

Evidently, the triple flip or CE transition rate ($\omega_{CE}$) is intricately dependent on the strength of the electron-electron ($e-e$) and electron-nucleus ($e-n$) couplings. In the case of a biradical with constant $e-e$ coupling ($\omega_{ee}$), when we extend $r_{en}$, the $e-n$ coupling ($\omega_{en}$) diminishes. Consequently, the efficiency of CE transfer also decreases. Taking this CE Hamiltonian into consideration, we explore the feasibility of tailoring the spatial arrangement of two electron spins, with the aim of facilitating  CE DNP transfer for nuclear spins situated at varying distances.

\begin{figure}[H]
    \centering
    \includegraphics[width=1.0\textwidth]{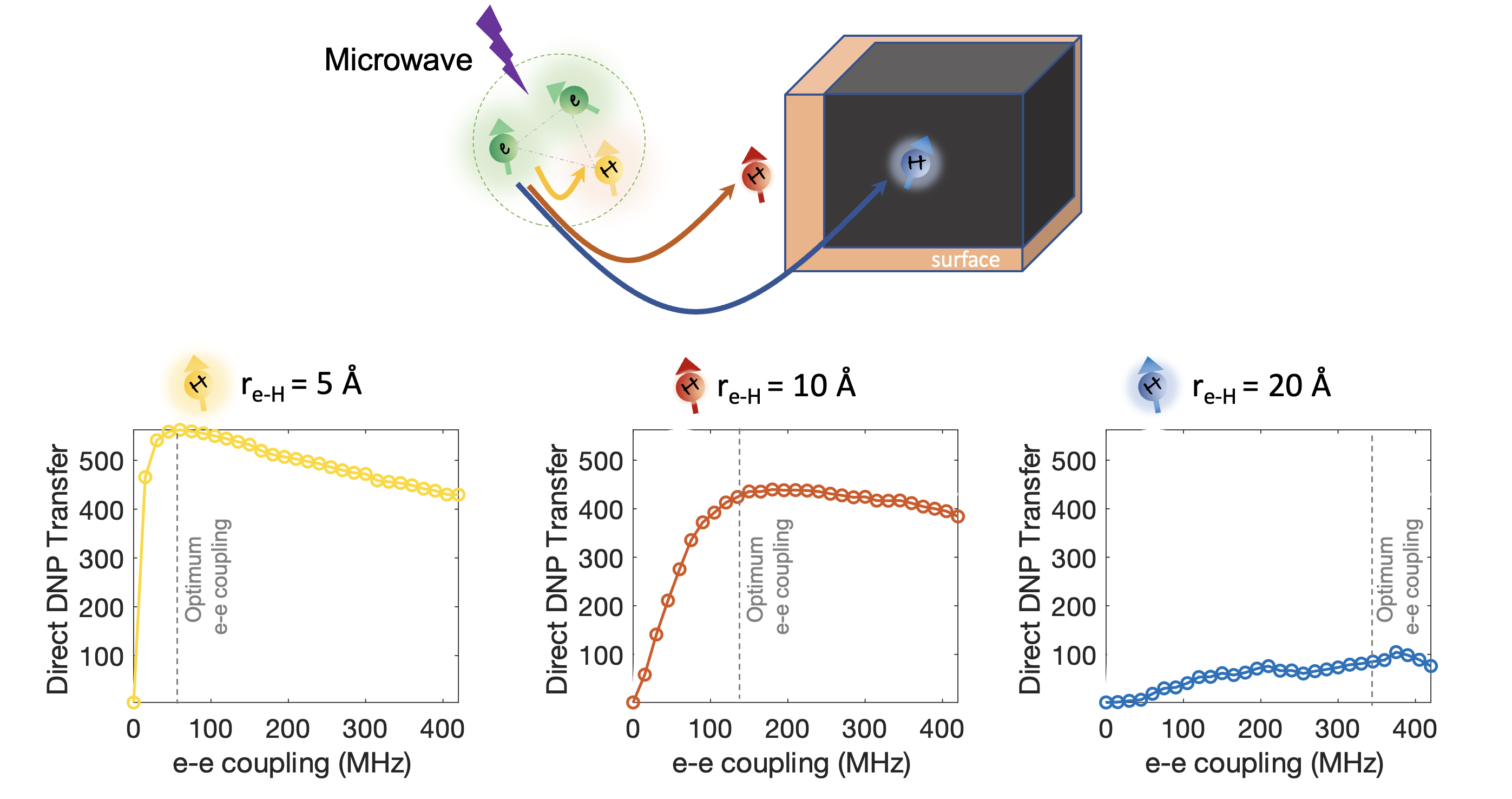}
    \caption{Quantum mechanically simulated direct CE DNP transfer as a function of $e-e$ coupling for three scenarios with e-H distances fixed to: 5 $\AA$, 10 $\AA$, and 20 $\AA$. Stronger $e-e$ coupling is required for maximum DNP enhancement over longer e-H distances. The simulations were performed using a $eeH$ spin system mimic the experimental condition of 7 T filed and 10 kHz spinning. Refer to the `Methods' section for detailed spin system information.}
    \label{fig3}
\end{figure}

In our study on CE DNP transfer, we explored three scenarios involving polarization transfer to a $H$ at distances of 5 $\AA$ (short), 10 $\AA$ (intermediate), and 20 $\AA$ (long). Polarization transfer to a shorter distance illustrates transfer within the biradical, while intermediate distances reflect transfer to solvent nuclei, and longer ranges represent direct transfer to the target molecule. We maintained constant spin parameters, including relaxation rate constants, microwave power, and frequency. The only variable under changed was the $e-e$ coupling strength, examining how changes in $r_{en}$ affected optimal $e-e$ coupling. Our findings reveal a compelling relationship between $e-e$ coupling and DNP enhancement for varying $e-H$ distances. At the shortest $e-H$ distance (5 $\AA$), we observed a maximum DNP enhancement of 550 with an $e-e$ coupling strength of $\approx$ 50 MHz. Extending the e-H distance to 10 $\AA$, we found a stronger $e-e$ coupling of $\approx$ 120 MHz was needed for a maximum DNP enhancement of 450. At the longest e-H distance (20 $\AA$), an even stronger $e-e$ coupling exceeding 300 MHz was required for a maximum DNP enhancement of $\approx$ 100. These simulations underscore a crucial trend: the need to tailor the $e-e$ coupling strength depending on the transfer scenario. Direct DNP transfer over longer e-H distances necessitates significantly stronger coupling than conventional DNP-mediated transfers to short distances and subsequent spin diffusion. Our analysis also indicates that in the current experimental setup (microwave power much weaker than e-e coupling), SE DNP cannot achieve long-range transfer. This insight carries substantial implications for understanding and optimizing DNP processes, providing valuable guidance for enhancing DNP efficiency in spin systems with variable $e-H$ distances.

\subsection{Time dependent effects from Magic Angle Spinning }

\begin{figure}[H]
    \centering
    \includegraphics[width=0.75\textwidth]{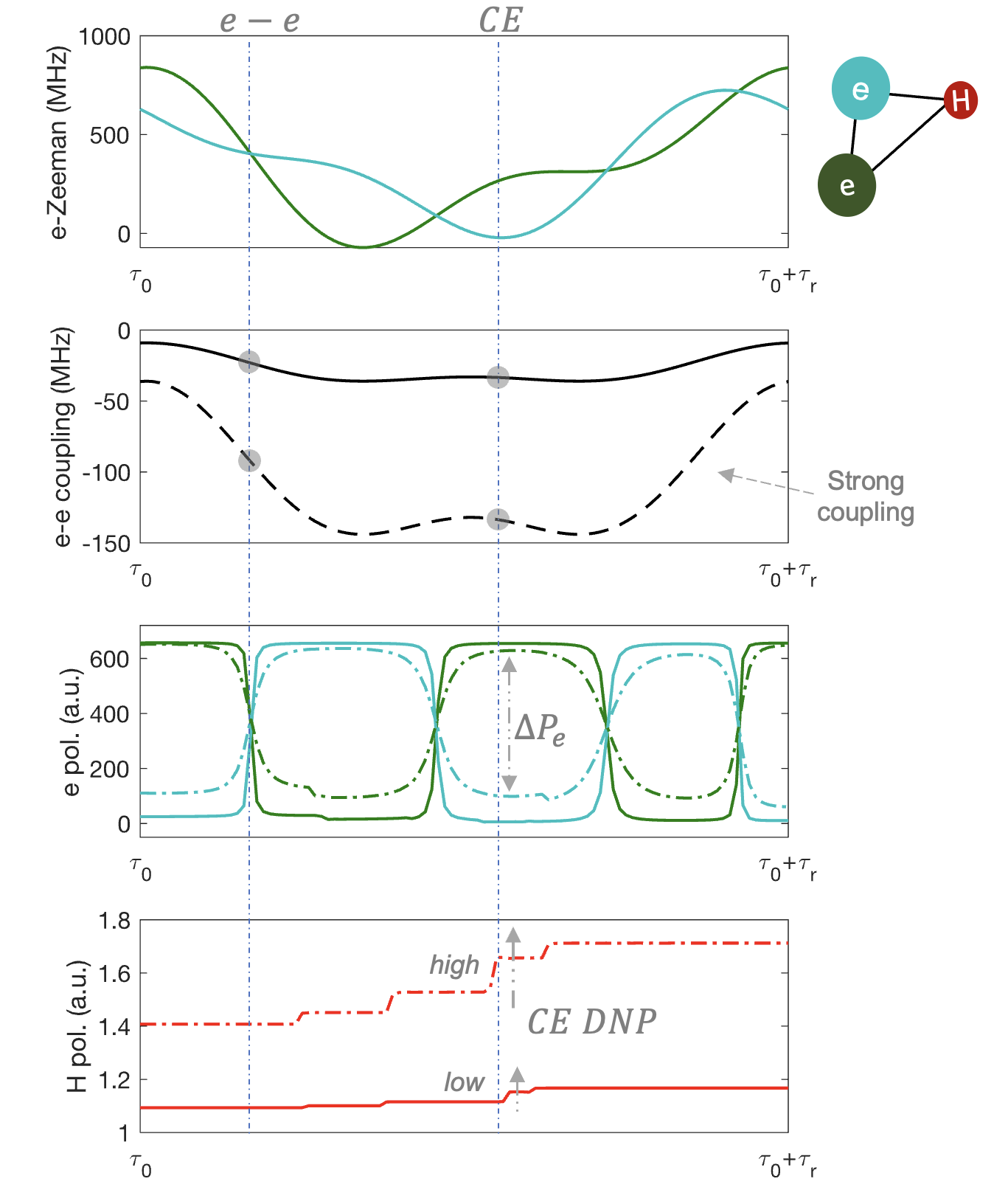}
    \caption{The Zeeman energy trajectories of two electrons for a single orientation of bis-nitroxide radical under MAS are displayed in Panel (i). Vertical lines indicate $e-e$ and CE rotor events. Electron-electron coupling trajectories are depicted in Panel (ii). Polarization of electron and nuclear spins under MAS-DNP condition for intermediate (solid lines) and strong (dashed lines) $e-e$ couplings are shown in Panels (iii) and (iv), respectively. When the $e-e$ coupling is too strong, it minimizes the polarization difference between electrons (Panel $iii$) but increases the adiabaticity of the CE events (Panel $iv$).}
    \label{fig4}
\end{figure}

DNP mechanisms vary significantly between static and MAS settings. In MAS DNP involving a biradical, three critical time-dependent interactions emerge: 1-spin flip (induced by microwaves, saturating electron spins), 2-spin $e-e$ flip-flop (leading to electron polarization exchange through coupling), and 3-spin triple-flip (leading to CE DNP transfer) transitions. These interactions are periodic and temporally separated due to sample rotation, as explained in recent literature \cite{thurber2012theory, mentink2015theoretical,gkoura2023primer}. For a comprehensive understanding of CE DNP's microscopic intricacies, we employ the Landau Zener model \cite{thurber2012theory, MENTINKVIGIER2012}. Notably, nitroxide radicals' g-anisotropy at high magnetic fields induces time-dependent energy levels for electron spins (Fig. \ref{fig4}, panel-i), resulting in dynamic resonance or \textit{rotor events} during MAS. One of the $e-e$ and $CE$ rotor events are highlighted for visualization. During these rotor events, Zeeman energies of the electrons converge and diverge from one of the resonance conditions, leading to real-time rotation and mixing of eigenstates, and hence causing transfer of population between states. This phenomenon is known as level anti-crossings (LACs) \cite{ivanov2021floquet}. The probability of the population transfer depends on the interplay between the rate of energy change ($\frac{d E_0}{dt}$) and the magnitude of perturbation at the LAC. Table \ref{tab:0} lists three distinct LAC types and their corresponding Landau Zener parameters.

To elucidate our numerically simulated findings from Fig. \ref{fig3}, we microscopically analyze DNP of a single, optimally oriented, biradical in a spinning rotor. Our observations elucidated the need for stronger $e-e$ coupling in CE DNP transfers to more distant nuclei for long $e-H$ distances. We present a microscopic energy and spin polarization trajectory analysis at an $e-H$ distance of 10 $\AA$ for two distinct $e-e$ couplings: (1) an intermediate $e-e$ coupling of 60 MHz, ideal for conventional DNP (short-range transfer), and (2) a stronger $e-e$ coupling of 180 MHz optimal for transfer at 10 $\AA$. It is important to mention that the $e-e$ coupling ($\omega_{ee}$) encompasses both exchange (J) and dipolar (D) contributions and therefore modulates under MAS due to dipolar anisotropy. The modulation is within the range: $[\frac{-D}{2}+J,D+J]$, as depicted in Fig. \ref{fig4}, panel-ii. The Zeeman energies of the two electron spins, calculated in the $\mu$w rotating frame, remain the same in both cases 1 and 2, as demonstrated in panel-i.

From Table 1, it is evident that the $e-e$ coupling plays a major factor in both $e-e$ and $CE$ rotor events. An optimal $e-e$ coupling facilitates adiabatic polarization exchange between the electron spins during e-e rotor events, generating a characteristic shape resembling two intersecting sigmoids. This exchange of polarization between electrons is crucial for maintaining a substantial polarization difference ($\Delta P_{e}$) needed for large polarization transfer at the CE rotor event. A 60 MHz $e-e$ coupling is sufficient for achieving adiabatic polarization exchange between electrons, as seen in the electron polarization profile in Fig. \ref{fig6}, panel iii (solid lines).

Contrarily, the triple-flip transition of the CE rotor event has a weak transition moment integral, as it is a second-order perturbative effect that is scaled down by the nucleus's Larmor frequency (Equation \ref{eq1}). For $\omega_{ee}$= 60 MHz and $r_{eH}$ = 10 $\AA$ (i.e. $\omega_{eH}$ $\approx$ 0.079 MHz), $\omega_{CE}$ will be just 0.016 MHz at 7 T. This small $\omega_{CE}$ leads to a small $^1H$ spin polarization enahncement during the CE rotor event (solid line, panel iv). As the $e-n$ distance increases further, the adiabaticity of the triple-flip transition diminishes, rendering conventional biradicals ineffective for long-range transfer. However, the decline of $\omega_{CE}$ for long $e-n$ distance can be mitigated by enhancing the $e-e$ coupling strength to 180 MHz, increasing $\omega_{CE}$ to 0.048 MHz (panel iii, dashed line ). This cumulative effect of multiple CE events effectively boosts nucleus polarization.

\begin{figure}[H]
    \centering
    \includegraphics[width=0.60\textwidth]{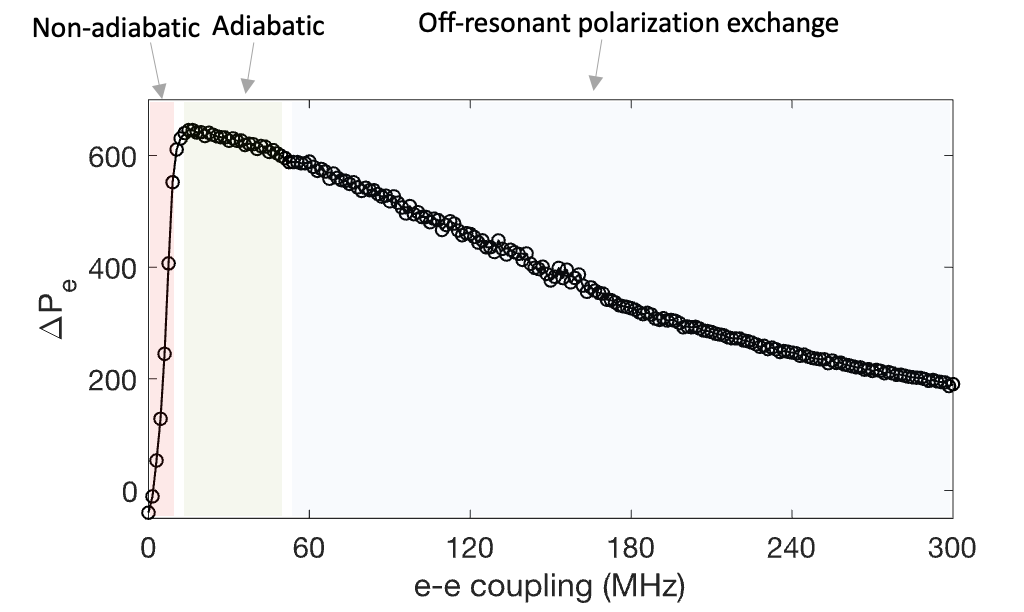}
    \caption{. The difference in polarization between electron spins ($\Delta P_e$) is dependent on the strength of the electron-electron coupling under MAS. When the coupling is too weak (red shade), the exchange of polarization is non-adiabatic, resulting in a small $\Delta P_e$. Optimal electron-electron coupling (green shade) leads to adiabatic exchange and the highest $\Delta P_e$. If the coupling is too strong (blue shade), $\Delta P_e$ is minimized due to off-resonance exchange effects.}
    \label{fig5}
\end{figure}

Notably, a large $e-e$ coupling enhances CE adiabaticity, promoting efficient and rapid CE transfer to coupled nuclear spins and facilitating polarization transfer to more distant nuclei. However, it is crucial to avoid excessively large $\omega_{ee}$. If the $e-e$ coupling becomes overly large, surpassing adiabaticity requirements, an undesirable phenomenon emerges in the electron polarization profiles. In such cases, the slope of the sigmoid-shaped polarization profile diminishes, resulting from polarization exchange even under off-resonant conditions due to significant perturbation. This minimizes the polarization difference between the electron spins, disrupting the high field approximation and increasing homogeneous coupling between them. This effect is further illustrated by mapping the polarization difference between electrons, $\Delta P_e$, at the CE rotor event as a function of $e-e$ coupling strength in Fig. \ref{fig5}. Clearly, both too weak and excessively strong coupling result in small $\Delta P_e$ due to non-adiabatic exchange and off-resonant polarization exchange caused by increased homogeneous coupling. Hence, achieving a balanced and customized e-e coupling based on the e-H distance for polarization transfer becomes a crucial consideration.

\begin{table*}
    \caption{Landau Zener parameters of LACs corresponding E$_1$, dE/dt and resonance conditions for two electron spins, $e_i$ and $e_j$.}    
    \centering
    \begin{tabular}{c|c|c|c|c}
            \hline
            Transition & Rotor-event & Perturbation (E$_1$) & dE/dt & Resonance Conditions \\  
            \hline
            1-spin & $\mu w$ &  $\omega_{1\mu w}$ & $\propto$ $g_{aniso}.B_0$, $\omega_r$  & $\omega_{0e_i}=\omega_{\mu w}$\\ 
            2-spin & $e-e$ & $\omega_{ee}$ &$\propto$ $g_{aniso}.B_0$, $\omega_r$ & $\omega_{0e_i}=\omega_{0e_j}$ \\
            3-spin & $CE$ &  $\frac{\omega_{e_ie_j}(\omega_{e_in}-\omega_{e_jn})}{\omega_{0n}}$ &$\propto$ $g_{aniso}.B_0$, $\omega_r$ & $\omega_{0e_i}-\omega_{0e_j} =\pm \omega_{0n}$ \\ 
            \hline
            \end{tabular}

    \label{tab:0}
\end{table*}

\begin{figure}[H]
    \centering
    \includegraphics[width=1.0\textwidth]{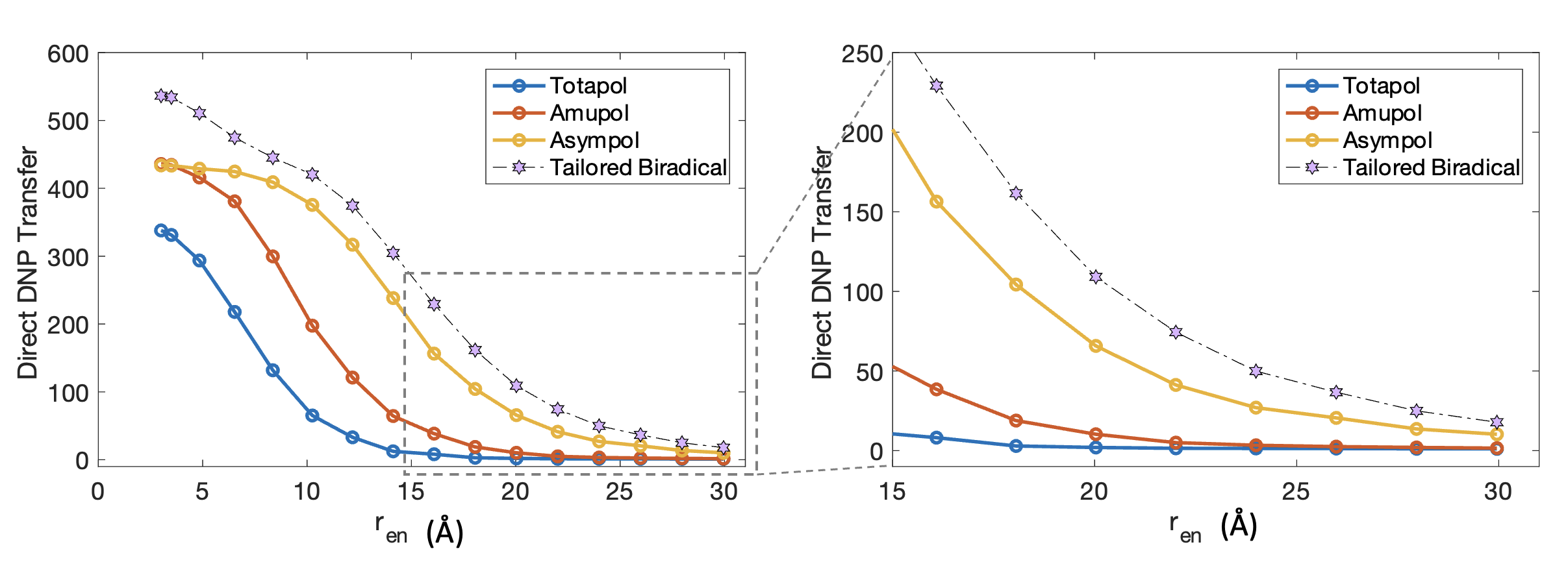}
    \caption{Quantum-Mechanically simulated direct CE DNP transfer as a function of $r_{en}$ for three Standard bisnitroxide polarizing agents. Here $n$ is a $H$ spin. Each data point in "$\star$" represents the performance of a tailored bisnitroxide polarizing agent at the specific $r_{en}$. The simulations were conducted to mimic experimental conditions at 7 T magnetic field strength and 10 kHz spinning frequency, using an $eeH$ spin system. Refer to the 'Methods' section for detailed information. . The panel on the right displays a zoomed in view of long-distance regime.}
    \label{fig6}
\end{figure}

\begin{figure}[H]
    \centering
    \includegraphics[width=0.6\textwidth]{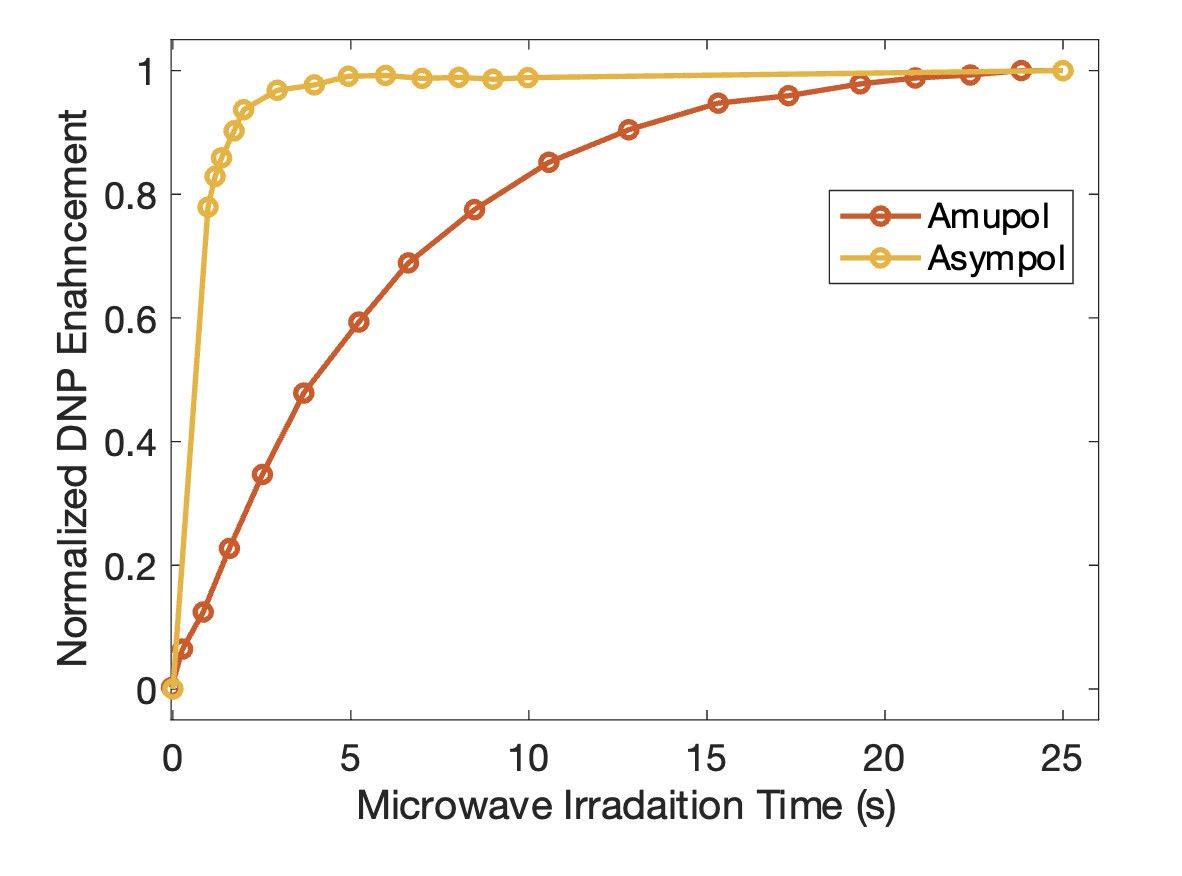}
    \caption{The experimentally obtianed buildup of $^1$H DNP of glycine using Amupol (red) and Asympol (yellow) biradicals dissolved in DNP juice matrix (d$_8$-glycerol:D$_2$O:H$_2$O). The experiment was conducted at 14 T, 10 kHz spinning and 100 kelvin condition.}
    \label{fig7}
\end{figure}

\section{Discussion and Outlook}

Many biradicals have been documented in the literature; however, a substantial number of these radicals do not perform optimally. Only the selected few have demonstrated efficient DNP performance, utilizing nuclear spin diffusion. Previous DNP studies have analyzed the combined effects of J-coupling and dipolar coupling, as well as their relative magnitudes compared to J and D \cite{mentink2018computationally,equbal2020balance}. It would be intriguing to analyze these radicals' performance in terms of their effectiveness in directly transferring polarization to a distant nucleus. Several parameters, such as relative g-tensors, relaxation rates, and electron-electron coupling strength, influence the radical's performance \cite{perras2017silico,mentink2017fast,mentink2019novo}. To simplify our analysis, we kept the relaxation rates constant, averaged over the relative g-tensor, and focused solely on the impact of inter-electron couplings on direct DNP transfer. In Fig. \ref{fig6}, we compare the CE DNP of three popular and commercially available polarizing agents: Totapol \cite{song2006totapol}, Amupol \cite{sauvee2013highly}, and Asympol \cite{Vigier2018}, at a 7 T magnetic field.

The coupling in Totapol is the smallest, and consequently, its performance deteriorates rapidly as the $e-H$ distance increases. Amupol offers better performance than Totapol but exhibits low transfer efficiency at longer distances, indicating that Totapol and Amupol are efficient when an effective spin diffusion network connects the biradicals to the solvent and subsequently to the target molecule. Direct polarization transfer in these two cases is inefficient. This inefficiency is exacerbated at higher magnetic fields and faster spinning rates than at the present conditions of 7 T and 10 kHz, respectively \cite{equbal2020balance,gkoura2023primer}.
Fig. \ref{fig6} also shows that among all commercially available bis-nitroxides, Asympol provides the best performance for long-range transfers. We also compare the performance of these bis-nitroxides with tailored biradicals, where the electron-electron coupling is optimized for each $e-H$ distances. Numerical simulations indicate significant potential for enhancing long-range DNP capabilities. At an $e-H$ distance of 15 $\AA$, a tailored biradical can achieve an enhancement of 240, compared to 140 obtained with Asympol at 7 T. Moreover, at an $e-H$ distance of 20 $\AA$, an enhancement of 110 can be achieved, compared to 60 obtained with Asympol at 7 T.

Asympol can transfer polarization over longer distances more effectively due to its strong electron-electron coupling of 200 MHz \cite{Vigier2018,equbal2020balance}. This feature facilitates direct transfer to the solvent molecule just outside the biradical, resulting in a faster DNP buildup. In Fig. \ref{fig7}, we compare the experimental buildup rates of Amupol and Asympol at 14 T, 100 K, and 10 kHz spinning, dissolved in a glycerol-water glassing matrix. At a concentration of 10 mM, Asympol exhibits a buildup rate of 0.9 s, whereas Amupol yields 6.8 s under the same experimental conditions. However, the experimentally obtained rate is a composite of contributions from different transfer pathways. A more controlled experiment is currently in progress and will be presented later.

In conventional spin diffusion-based mechanisms, the presence of nearby protons is indispensable. Proton-induced relaxation significantly impacts the efficiency of DNP transfer or sensing, underscoring its importance in developing effective DNP systems. In cases where DNP transfer relies on a polarizing agent via spin diffusion, the presence of strongly coupled protons becomes pivotal, as recently demonstrated by Venkatesh et al and Perras et al. in separate studies \cite{venkatesh2023deuterated, perras2020full}.  Incorporating a protonated biradical that establishes a robust proton network to facilitate efficient spin diffusion becomes essential in such scenarios.

Conversely, when DNP transfer occurs independently of the polarizing agent, an alternative strategy can be more advantageous. Deuteration of the molecular structure can emerge as a favorable approach in this context, as it extends the relaxation time constant for electrons \cite{canarie2020quantitative}. By mitigating the influence of proton-induced relaxation, deuteration significantly increases T$_{1e}$, presenting a valuable tactic to optimize DNP performance in systems where direct transfer outside the polarizing agent plays a predominant role. For efficient direct DNP transfer to distant target nuclei, it is also proposed to deuterate nearby hydrogen ($^1$H) spins around the electron spins. However, this effect requires experimental validation.

\section{Conclusion}

In this paper, we explore the effect of electron-electron interactions on the distance dependence of direct DNP transfer using the Cross Effect mechanism under MAS. We introduce a novel biradical design principle that extends the reach of DNP to more distant nuclei. We suggest the use of tailored biradicals for efficient transfer to short, intermediate and long-range $e-n$ distances. Notably, a bis-nitroxide optimum for short range transfer is not optimum for long range transfer and vice versa. For long range transfer, we propose the use of biradicals with strong $e-e$ coupling to expand the range of CE DNP. This could help to overcome the restrictions of conventional SE DNP and CE DNP methods and improve the applicability of DNP in solid-state NMR spectroscopy, particularly in cases where spin diffusion is inactive or direct sensing of nuclei using electron spin interactions is desired. By using a strongly coupled electron spin system, it is possible to unlock the full potential of CE DNP for efficient long-range polarization transfer, creating new possibilities for studying complex systems with higher sensitivity. For instance, a strongly coupled biradical could be used to polarize $^{19}$F spins labels in a biological sample even without the need of a fluorinated biradical or fluorinated solvent. Additionally, this could enable direct polarization transfer to the bulk/core of a material, as opposed to DNP-sens which selectively polarizes the surface \cite{Lesage2010}. Many radicals have also been proposed, some of which exhibited low performance for conventional spin diffusion based DNP \cite{zhai2018diastereoisomers}. However, these radicals can be utilized for their long range transfer, which is something under investigation in our lab and will be a subject of future discussion.

\section{Methods}

Cross Effect DNP is a complex phenomenon influenced by various underlying parameters. In experimental settings, the rational design of biradicals with diverse $e-e$ coupling, while maintaining constant spin parameters, proves challenging. Hence, we employed advanced quantum mechanical simulations within the Liouville space framework to investigate remote nuclear sensing via cross-effect DNP under magic angle spinning, employing the SPINEVOLUTION simulation package. Our model system consisted of two electrons ($e_1$ and $e_2$) and a proton ($^1$H) with preset g-tensor values (gx=2.0097, gy=2.0065, gz=2.0024), resembling a nitroxide radical. In practical scenarios, the relative molecular orientation of spin labels exhibits flexibility, which we accounted for by varying and averaging the relative g-tensor of $e_2$ with respect to $e_1$ using Euler angles ($\alpha$, $\beta$, $\gamma$) in the ($z-y-z$) convention. The magnetic field, spinning frequency, microwave irradiation frequency, microwave power, and temperature were maintained at 7T, 10 kHz, the peak of ZQ DNP, 800 kHz, and 100K, respectively, unless otherwise specified. We consistently utilized a protracted DNP buildup time (xx seconds) to achieve steady-state nuclear spin polarization. To ensure distinct hyperfine couplings for $e$ spins in all orientations, we selected the orientation of the $e_1-H$ hyperfine couplings carefully, allowing variations in hyperfine coupling or e-n distance magnitude while preserving orientation. Nuclear spin-lattice relaxation time (T$_{1H}$) remained at 2 seconds, while electron spin-lattice relaxation (T$_{1e}$) and spin-spin relaxation (T$_{2e}$) constants were set at 2 milliseconds and 10 microseconds, respectively, based on experimentally determined relaxation rates under similar conditions.

\subsection*{Acknowledgments}
The authors would like to thank New York University Abu Dhabi (NYUAD) for the financial support of this work, Core Technology Platforms and High Performance Computing facilities of New York University Abu Dhabi for facilitating experimental and theoretical DNP research. The authors would also like to thank Waqas Zia for the support with HPC. Contribution from AJ and AE was supported by Tamkeen under the NYU Abu Dhabi Research Institute grant CG008. We would like to thank Prof. Songi Han (UCSB) and Prof. Anne Lesage (CNRS) for fruitful discussions on DNP mechanisms.

\bibliographystyle{unsrt}
\bibliography{sample}


\end{document}